\documentclass[12pt]{article}
\usepackage{amsmath}
\usepackage{amssymb}
\usepackage{epsfig}
\title{Firm dynamics in a closed, conserved economy: 
A model of size distribution of employment and related statistics}
\author{Anindya S. Chakrabarti\\
Economics Department, 270 Bay State Road, MA, USA\\
Email id: {\it aschakrabarti@gmail.com}
}
\begin{document}

\maketitle

\begin{abstract}
\noindent 
\noindent We address the issue of the distribution of firm size. To this end we propose a model of firms in a closed,
conserved economy populated with zero-intelligence agents who continuously move from one firm to another. We then
analyze the size distribution and related statistics obtained from the model. Our ultimate goal is to reproduce the
well known statistical features obtained from the panel study of the firms i.e., the power law in size (in terms of income and/or
employment), the Laplace distribution in the growth rates and the slowly declining standard deviation of the growth rates
conditional on the firm size. First, we show that the model generalizes the usual kinetic 
exchange models with binary interaction to interactions between an arbitrary number 
of agents. When the number of interacting agents is in the order of the system 
itself, it is possible to decouple the model. We provide some exact results 
on the distributions. Our model easily reproduces the power law.  
The fluctuations in the growth rate falls with increasing size following a power law
(with an exponent 1 whereas the data suggests that the exponent is around 1/6). 
However, the distribution of the difference of the firm-size in this model
has Laplace distribution whereas the real data suggests
that the difference of the log sizes has the same distribution.
\end{abstract}

\section{Introduction} 
\noindent 
{
\noindent It is long known that the size distribution of the firms has a long tail which is remarkably robust \cite{gibrat}. A few very large firms can operate
side by side with a large number of small firms. Ref. \cite{axtell;01} presents clear evidence
that the distribution can be characterized very well by a power law and regarding the stability of the law, as the same reference puts it, this feature 
has survived changes in the political, regulatory and social regimes; the last one being caused by the demographic changes in the work force due to the
influx of women in the labor force. Also, numerous innovations and technological changes in the production process were unable to affect it. Lastly,
firm mergers, acquisitions, death and birth of firms did not affect this feature. Ref. \cite{stanley;96} studies the panel data on the firm dynamics
(all publicly traded US manufacturing firms in the time span 1975-1991)
and concludes that the growth rates show two more significant features. One, 
the distribution of the growth rates of the firms has an exponential form and the standard deviation
of the growth rates of the firms fall with increasing firm sizes following yet another power law.

\medskip

\noindent The above finding indicates that the statistical features for the firm growth process
are independent of
microeconomic decision-making processes (at least, to a first approximation) like why people choose to leave their job etc. 
Hence, we do not indulge in providing any microeconomic foundation for the firm dynamics.
However, the rate at which the firms gain and lose workers is of interest to us. This rate is called
the turnover rate in the economics literature. An alternative but closely related 
interpretation of the turnover rate
is that it measures how long the employees stay in their respective jobs. It may be noted that
the rates of hiring and separation for developed economies are very high. For example, in USA (2009-2011),
the average total seasonally adjusted annual hiring rate and separation rate was around 38-40\% 
(see Ref. \cite{data-turnover}). Hence, the turnover rates (interpreted as the average length 
of employment) may be very low. We intend to show that the turnover rates play a crucial role
in the firm size distribution and related issues. One important aspect of job separation and worker
hiring is that the process follows the rule of local conservation. If one worker goes from one firm to another
then the total workforce remains unchanged but the workers' distribution across the firms change.
Since the workers at any given year (or quarter) move around in a very large number of firms,
we model this process as a repeated interaction between a large number of agents (firms) which
exchanges a finite amount of (number of) workers between themselves.  Clearly, the idea of the
kinetic exchange model is suitable for this purpose.

\noindent In this paper, we present a model of firms in a closed, conserved economy populated with zero-intelligence agents.
The firms are modeled as collections of agents who
continuously move from one firm to another. But on an aggregate level, there is no fluctuation. 
The firm's size is solely determined by the number of agents working in the firm. We also assume that time is discrete. 
\noindent The basic idea of the model is that each period there is a group of agents in each firm who wants to move to another firm
with some personal motives (like utility maximization due to wage increase etc.) which we do not model here.
We call the turnover rate $\lambda$. Hence, $(1-\lambda)$ fraction of each firm's workforce
would want to move out of the respective firms. 
Each period there will be a pool of such agents who wants to shift from one firm to another. Some of them will find a new job (that is they will move to new firms)
while the rest has to continue in their earlier position. There is no unemployment in the model. This process is repeated until the distribution of workers
settle down and this distribution will be the size distribution of the firms. We shall show that
this model is closely related to the kinetic exchange models of markets and in fact, it generalizes the usual binary trading (collision) model
to interactions between an arbitrary number of agents (firms in this case). 
The exact distributions in some cases, will be provided.
Subsequently, we shall study a modification of the basic process which leads to a
Power law in the size distribution of the firms. Then we show that the standard deviation
decreases with increasing size following yet another power law and we study the 
corresponding distributions of the growth rates.

\medskip

\noindent The related literature is varied and vast. Ref. \cite{gibrat} is probably the first systematic treatment of the subject. To model the growth of a firm
it suggested a stochastic process which essentially states that the growth rate of the firm is independent of its size. This prediction and its result that
the distribution of the firm size would be log-normal, was found to be approximately correct \cite{stanley;95}. 
However, there are evidences that the formulation was not entirely correct. Ref. \cite{singh} first observed that
the standard deviation in the growth rate falls as the firm size increases. This finding is supported in later studies as well (see Ref. \cite{stanley;96, evans, davis, hall, duanne}).
Ref. \cite{stanley;97a} collects data for US manufacturing firms (1974-1993) which support their earlier
finding in \cite{stanley;96}. The more important part of this study is that they showed that the proposed statistical features are robust to firm birth and death
due to merger or bankruptcy. For an overview, see Ref. \cite{econphysfirms} and references therein. See also Ref. \cite{ishikawa}
for a detailed analysis of the Gibrat's law, Pareto index and Pareto law. The data set used is mainly the
panel data of the Japanese firms. 
On the theoretical ground, Ref. \cite{simon} contains study on stochastic properties of the dynamics of firm growth. Ref. \cite{stanley;97b}
presents a model of hierarchical organizations to explain the observed regularities. 
See Ref. \cite{aoyama, mizuno} for a separate theoretical approaches to the dynamics 
of company growth. 
See Ref. \cite{acbkc;07, chakraborti;10, yako-rosser;09, chakrabartis;10} for detailed
discussion of the kinetic exchange models of markets. Lastly, Ref . \cite{jovanovic} provides theories of job matching and turnover.

\medskip

This paper is organized as follows.
In section \ref{sec: basic}, we propose the basic model and derive the exact distribution of workers in the firms. In the
next section, we modify the model which leads to the power law distribution of the firm sizes. 
In section \ref{sec: growth}, we show the distributions of the growth rates of the firms in this economy
and compare them with real data. 
Then follows a summary and in the appendix, we have presented short discussions 
on how the kinetic exchange models
are related to the generalized Lotka-Volterra equations and also, how can we
apply the model stated in this paper directly to model the income/wealth
distributions.
. 
}
\section{The model with constant turnover rate, $\lambda$} 
\label{sec: basic}
{
\noindent We assume that time is discrete. The economy consists of an array of $N$ firms which can absorb any amount of workers that come to it.
The workers are treated as a continuous variable (infinitely divisible).
At the very beginning of the process, all firms have exactly one unit of workers (more formally, the measure of workers is one for each firm).
The fraction of workers that decides to stay back in their firm (which we interpret as the turnover rate), 
is denoted by $\lambda$ which may vary between the firms. For the time being,
we treat them as given and constant across the firms. This treatment is pioneered by Ref. \cite{anirbanc;00}
in the context of modeling income/wealth distributions.
As we said earlier, the firm's size is just the measure of workers working in the firm. 
There are other indicators of the firm size as well e.g.
quantity of goods produced, sales, cost of goods sold, assets or value of the properties. But we note that not all firms 
produce the same, identical goods. The inputs also differ very much. To consider capital holding (or the value of assets), that
is not always easy to measure (large fluctuations happen in the stock market in short spans of time adding or wiping huge amounts from the value).
Hence, it is easier to use the size of the workforce as a proxy for the firm size. We denote the firm size of the $i$-th firm (that is the work force) by $w_i$ ($i~\le ~N$ where $N$ is the number of firms). Also, suppose that the number of firms from which the workers are leaving and moving into, is $n$. At each time point $(1-\lambda)$ fraction of the workforce of those $n$ firms wants to leave. So there would be a total pool of workers that wants to change their workplace. Next, this pool of workers is randomly divided into those $n$ firms. Hence, the dynamics is given by the following set of equations,

\begin{eqnarray}
w_1(t+1) &=& \lambda w_1(t) +\epsilon_{1(t+1)}(1-\lambda){\sum^n w_j(t)} \nonumber\\
\ldots         \ldots \nonumber\\
w_i(t+1) &=& \lambda w_i(t) +\epsilon_{i(t+1)}(1-\lambda){\sum^n w_j(t)} \nonumber\\
\ldots        \ldots\nonumber\\
w_n(t+1) &=& \lambda w_n(t) +\epsilon_{n(t+1)}(1-\lambda){\sum^n w_j(t)} 
\label{constlambda}
\end{eqnarray}

\noindent such that $\sum^n_j \epsilon_{j(t)} ~=~1$ for all $t$. As is evident from above, this is a straight generalization of the usual kinetic exchange models (with $n=2$) that has primarily been used to study the income/wealth distribution models (see Ref. \cite{acbkc;07, yako-rosser;09, chakrabartis;10}). Regarding the notations, we use $t$ within 
the first bracket when referring to the endogenous variables like the size of the firm ( $w(t)$) 
and we use the same in subscript
when referring to the exogenous random variables (e.g., $\epsilon_t$). Similarly,
with a slight abuse of notation, we denote
the probability density function ($pdf$) of the exogenous random variable $x$ by $f(x)$. However, for the
distributions of the endogenous random variables (for example, the distribution of the firm size or of the
growth-rate), we use $P(.)$ ($P(w)$ and $P(g)$ resp.).

\medskip

\noindent {\bf Construction of $\epsilon$}

\medskip
{
\noindent Here, we consider a few constraints on $\epsilon$. 
\begin{enumerate}
\item The sum of all $\epsilon_i$s has to be equal to one.
\item The expectation, $E( \epsilon_i $) = 1/$n$ for all $i$ and the distributions of all $\epsilon_i$ are identical.
\item If $n=2$, $\epsilon_i \sim$ uniform[0, 1]. We impose this constraint so that at the lower limit of $n$, we
get back the usual CC-CCM models (see Ref. \cite{acbkc;07}).
\end{enumerate}

\noindent Formally, the problem then boils down to that of sampling uniformly from the unite simplex (see Ref. \cite{onn-smith}). We follow the standard algorithm and below we derive the distribution of $\epsilon$.
\begin{enumerate}

 \item Create a vector of independent random variables drawn from uniform distribution over 
[0, 1], $\xi_1$, $\xi_2$, \ldots , $\xi_n$. 

\item Take logarithm of all the elements of the vector and multiply the elements by -1.

\item Divide each element by the sum of all the elements. Call the $i$-th result $\epsilon_i$ for all $i$.

\end{enumerate}

\noindent We derive the probability density function of the $\epsilon_i$ below. Consider $\epsilon_1$ first for simplicity. The probability that $\epsilon_1$ is less than some $\theta$ is

\begin{eqnarray*}
Prob. (\epsilon_1<\theta) &=& Prob. (-\ln\xi_1<-\theta \ln (\xi_1\xi_2\ldots\xi_n))\\
               &=&1-Prob.(\xi_1<\xi_2^{\frac{\theta}{1-\theta}}\xi_3^{\frac{\theta}{1-\theta}}\ldots\xi_n^{\frac{\theta}{1-\theta}})\\
               &=& 1-[\int_0^1 \xi_2^{\frac{\theta}{1-\theta}} d\xi_2]^{n-1}  \hspace{1 cm}\mbox{(using independence)}\\
               &=& 1-(1-\theta)^{n-1}.
\end{eqnarray*}

\noindent This is true for all $\epsilon_i$. Therefore, the {\it pdf} of $\epsilon_i$ is
\begin{equation}
f(\epsilon_i)=(n-1)(1-\epsilon_i)^{n-2},
\label{epsilon-beta}
\end{equation}
\noindent that is, $\epsilon$ has a beta pdf with parameters 1 and $n-1$. Clearly, when $n=2$ the distribution of $\epsilon$ is uniform[0, 1] as expected.

\begin{figure}
\begin{center}
\noindent \includegraphics[clip,width= 12cm, angle = 0]
{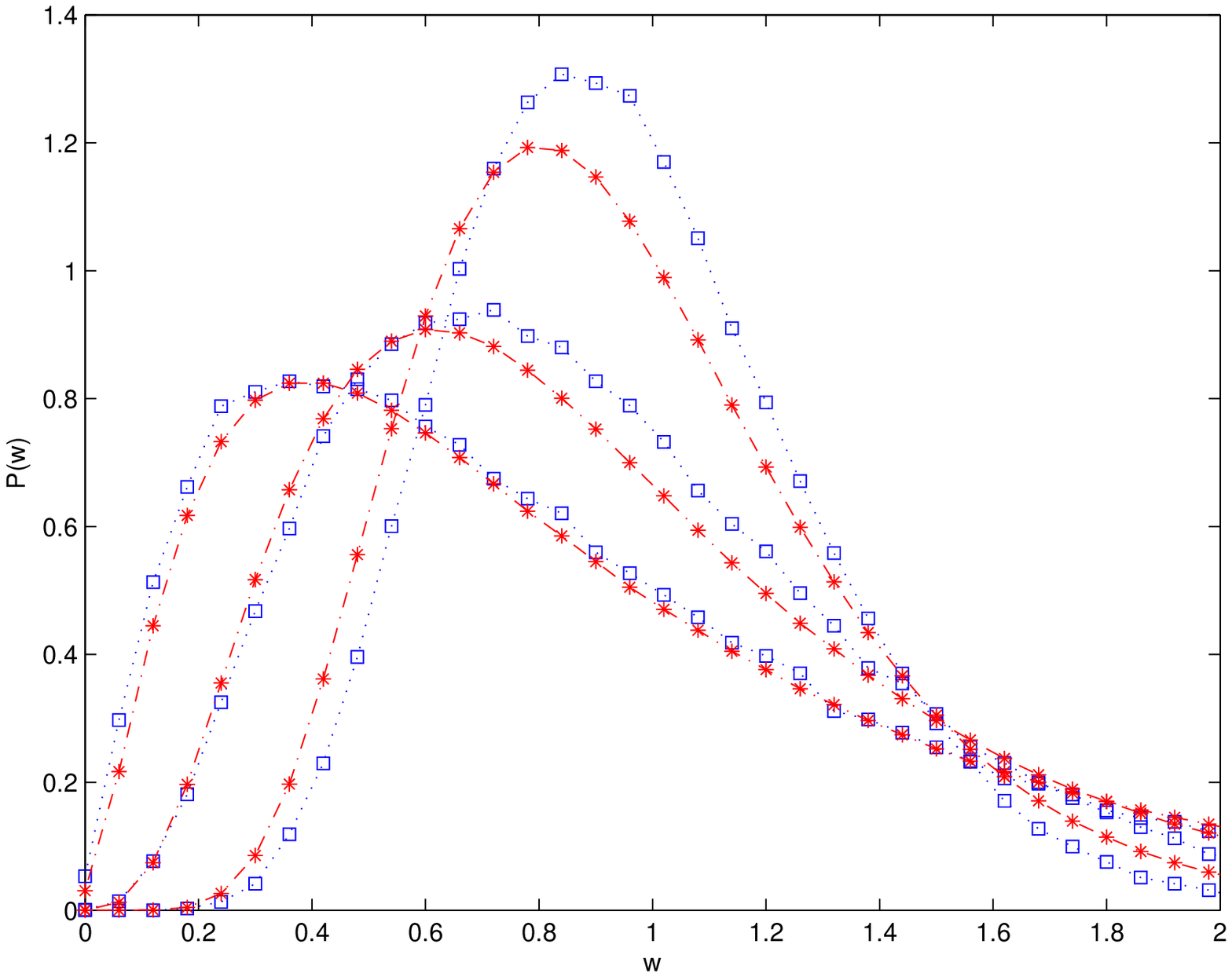}
\caption\protect{Workers' distribution across the firms: comparisons between binary interaction i.e., usual
kinetic exchange model (blue squares; see Ref. \cite{acbkc;07}) 
and $N$-ary interaction model (red stars). Three cases are shown
above, viz., $\lambda=1/4$, $\lambda=2/4$,
$\lambda=3/4$.
All simulations are done
for $O(10^4)$ time steps with 1000 agents and averaged over $O(10^3)$ time steps.
Note that in Sec. \ref{subsec: ss} we have derived that for $\lambda=0$, both curves are identical. 
Discrepancies appear for $\lambda>0$ as is apparent in the above diagram.
}
\label{CC-ASC}
\end{center}
\end{figure}
}

\subsection{Reduced form of the model}
\label{subsec: reduced}
{
\noindent First, we note that the solution to the usual kinetic exchange model with binary interaction 
is not known yet (see Ref. \cite{acbkc;07, chakraborti;10}). The resultant distribution is approximated by gamma probability distribution function \cite{chakraborti;04}. But moment considerations show that the distribution does not have a
gamma form (see Ref. \cite{richmond;05}). Here, we derive an exact result of the case where the number of interacting firms is in the order of the system size $N$ i.e., we consider the case where $2<<n\le N$.

\medskip

\noindent Note that if $n$ is of the order of $N$,  
$\sum^n_j w_j$ is well approximated by $n$ (recall that $E (w_j) =1$ for all $j$). To make sure, note that
$\sum^N_j w_j$= $N$ by the structure of the model. For exactness, we shall assume that all firms interact at
every step, i.e., $n=N$. Then the system of equation becomes

\begin{eqnarray}
w_1(t+1) &=& \lambda w_1(t) +\epsilon_{1(t+1)}(1-\lambda)N \nonumber\\
\ldots         \ldots \nonumber\\
w_i(t+1) &=& \lambda w_i(t) +\epsilon_{i(t+1)}(1-\lambda)N \nonumber\\
\ldots         \ldots \nonumber\\
w_n(t+1) &=& \lambda w_n(t) +\epsilon_{n(t+1)}(1-\lambda)N
\label{constlambda1}
\end{eqnarray}
\noindent with each $\epsilon_i$ having a beta distribution as has been found in 
Eqn. \ref{epsilon-beta} (see Construction of $\epsilon$ in Sec. \ref{sec: basic}). Note that in this form, we get rid of the effects of $w_j(t)$ in the
evolution equation of $w_i(t)$ for all $j\ne i$. One more simplification is possible.
Let $\mu=N(1-\lambda)\epsilon$ ignoring the subscripts. Given $N$, it is easy to verify that
the probability distribution of $\mu$ is 
\begin{equation}
f(\mu)=\frac{N-1}{N(1-\lambda)}\left(1-\frac{\mu}{N(1-\lambda)}\right)^{N-2}.
\end{equation}

\noindent Hence, for large $N$ we can approximate the distribution as the following,
\begin{equation}
\lim_{N\rightarrow\infty}f(\mu)\simeq\psi e^{-\psi \mu} \hspace{1 cm}
\mbox{where $\psi=\frac{1}{1-\lambda}$}.
\label{errdistr}
\end{equation}

\noindent Therefore, the system reduces to
\begin{eqnarray}
w_1(t+1) &=& \lambda w_1(t) +\mu_{1(t+1)} \nonumber\\
\ldots         \ldots \nonumber\\
w_i(t+1) &=& \lambda w_i(t) +\mu_{2(t+1)} \nonumber\\
\ldots         \ldots \nonumber\\
w_N(t+1) &=& \lambda w_N(t) +\mu_{N(t+1)},
\label{constlambda2}
\end{eqnarray}

\noindent which is a system of autoregressive type equations with the distribution of errors
($\mu$) given by Eqn. \ref{errdistr}.
}

\subsection{Steady state distributions}
\label{subsec: ss}
\subsubsection{For $\lambda$ = 0}
\label{subsubsec: ss0}
{
In the above section (Sec. \ref{subsec: reduced}), we have derived the reduced form equations. Below, 
we find their solutions. One noteworthy feature is that with $\lambda$ = 0, the system further reduces to

\begin{equation}
w_i(t) = N\epsilon_{it} \hspace{1 cm}\mbox{for all $i$.}\\
\end{equation}

\noindent As we showed in the above section, the steady state distribution would be exponential. Note that 
the result is identical to the case where the interaction is binary. Also, we can provide another proof by conjecture. Let us rewrite the system as
 \begin{equation}
w_i(t) = \epsilon_{it} \left({\sum^N_j w_j(t-1)}\right) \hspace{1 cm}\mbox{for all $i$.}\\
\end{equation}

\noindent Let us conjecture that the steady state distribution is exponential. More precisely, let 
$f(w_j)=exp(-w_j)$ for 
all $j$. Clearly, $\sum_j^N w_j(t-1)$ has a gamma pdf with parameters 1 and $N$. Recall that
$\epsilon$ has a beta pdf with parameters 1 and $N-1$. Therefore the distribution of 
their product is again exponential (see Thm. 2.3 in Ref. \cite{veleva}) confirming our conjecture.

\subsubsection{With positive $\lambda$}
\label{subsubsec: ss}
{
\noindent However, the above result (the equivalence between the distributions generated by binary interactions and N-ary interactions) does not hold in presence of positive $\lambda$. 
First, we discuss the discrepancies in the second moment. Then we move on to derive
the exact distribution.
\noindent We denote 
the central moment of order $\bar{n}>1$ of a variable $x$ as
$$E(x-E(x)^{\bar{n}}) = E(\sum_{l=0}^{\bar{n}}\left(\begin{array}{c}\bar{n}\\l\end{array}\right)
x^lE(-x)^{(\bar{n}-l)}).$$ 
For $\bar{n}=2$, $E(x-E(x)^{\bar{n}})$ corresponds to the variance of $x$ and is denoted by $V(x)$.
Since the system is conservative and the initial workforces (i.e., the firm sizes) were
unity for all firms, it is obvious that $E(w_i)$ would be unity.
So we can write the $n$-th moment of the distribution of size without subscript as
\begin{equation}
\label{moment}
E((w-1)^{\bar{n}}) = E(\sum_{l=0}^{\bar{n}}\left(\begin{array}{c}\bar{n}\\l\end{array}\right)(-w)^l). \end{equation}
We assume that $w_i$ and $w_j$ are independent variables (technically, they are not since the sum
of all $w_i$s is constant, $N$ in this case; but for large $N$ this is a good approximation ). It is easy to verify that with all firms interacting ($n=N$), the variance is given by
$$V(w)=\frac{(1-\lambda)}{(1+\lambda)}$$ \noindent whereas in the case of binary interaction
\cite{chakrabartis;10}
$$V(w)=\frac{(1-\lambda)}{(1+2\lambda)}.$$
 \noindent Note that for $\lambda=0$, variance is unity in both cases which is consistent with our derivation
that the distribution is the same (exponential) in both cases.

\noindent Let us write the system as
\begin{equation}
w(t+1) = \lambda w(t) +\mu_{t+1} \nonumber\\
\end{equation}
\noindent which can be rewritten with the lag operator $L$ as 
$(1-\lambda L)w(t) =\mu_t $
and hence,
\begin{equation}
w(t)=\mu_t+\lambda\mu_{t-1}+\lambda^2\mu_{t-2}+\lambda^3\mu_{t-3}+\ldots .
\end{equation}
\noindent Recall that (Eqn. \ref{errdistr})

\begin{equation}
f(\mu)\simeq\frac{1}{1-\lambda}e^{-{\frac{1}{1-\lambda}} \mu}.\nonumber 
\end{equation}

\noindent Therefore in the steady state, 
\begin{equation}
w=\tilde{\mu}_0+\tilde{\mu}_1+\tilde{\mu}_2+\tilde{\mu}_3+\ldots .
\end{equation}
\noindent where $\tilde{\mu}_j$ is distributed as
$$f(\tilde{\mu}_j)=\frac{1}{\lambda^j(1-\lambda)}e^{-\frac{\tilde{\mu}_j}{\lambda^j(1-\lambda )}}.$$
\noindent We can neglect the terms with high powers (more than say $k$) of $\lambda$. 
Then $w$ is essentially
 the sum of $k$ exponentially distributed random variables with different parameters. 
Note that the Laplace transformation $L(s)$ of $\mu_j$ is $\phi_j/(\phi_j+s)$ with 
$\phi_j=1/(\lambda^j(1-\lambda)).$
Since the $\mu_j$'s are $i.i.d.$, pdf of $w$ would be the convolution of the pdfs of the $k$ 
random variables. By
property of Laplace transformation, it can be verified that the distribution of $w$ would be
(see Ref. \cite{cordecki} for detailed discussions and different proofs)
\begin{equation}
f(w)=\sum_{i=1}^k \phi_i exp(-\phi_i w) \prod_{j=1,j\ne i}^k \left( \frac{\phi_j}{\phi_j-\phi_i} \right)
\end{equation}

\noindent with $\phi_i$ defined as $\phi_i=1/(\lambda^i(1-\lambda))$ (see figure \ref{CC-ASC}). 

}

}

\section{Distributed turnover rates, $\lambda_i \ne \lambda_j$ } 
\label{sec:distr lambda}
{
\noindent So far, we have considered only fixed $\lambda$. 
In this section we consider distributed $\lambda$ (i.e., the turnover rates differ across firms but they are fixed over time) following Ref. \cite{acbkc;07}. Specifically, we assume that the turnover rates are
uniformly distributed over the interval [0, 1] across the firms. 
The new system of equation is

\begin{eqnarray}
w_1(t+1) &=& \lambda_1 w_1(t) +\epsilon_{1(t+1)}{\sum^n_j (1-\lambda_j) w_j(t)} \nonumber\\
\ldots    \ldots      \nonumber\\
w_i(t+1) &=& \lambda_i w_i(t) +\epsilon_{i(t+1)}{\sum^n_j (1-\lambda_j) w_j(t)} \nonumber\\
\ldots       \ldots   \nonumber\\
w_n(t+1) &=& \lambda_n w_n(t) +\epsilon_{n(t+1)}{\sum^n_j (1-\lambda_j) w_j(t)} 
\label{distrlambda}
\end{eqnarray}

}
\begin{figure}
\begin{center}
\noindent \includegraphics[clip,width= 12cm, angle = 0]
{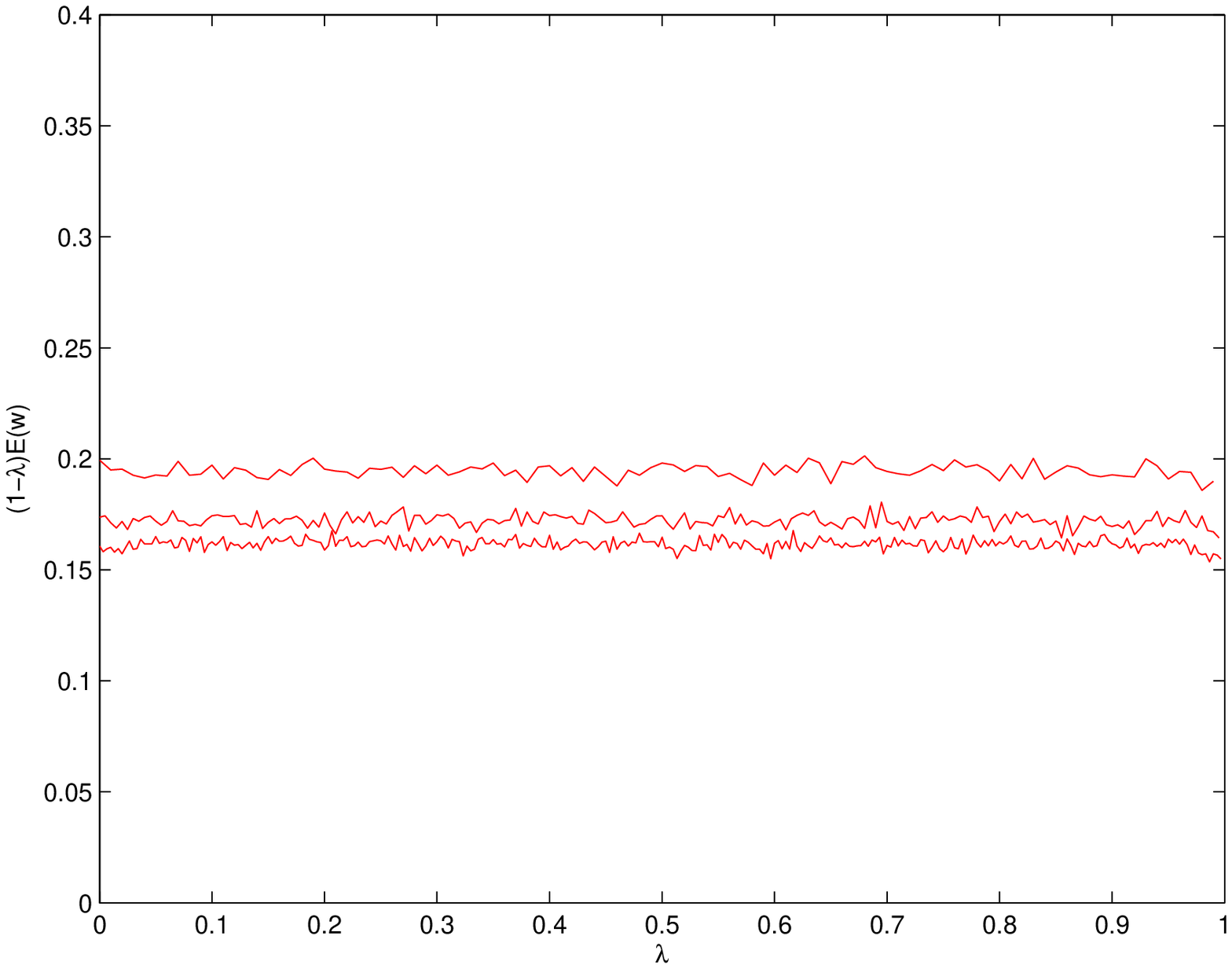}
\caption\protect{Finding the value of $C$ from Eqn.  \ref{distrlambda1}. We have considered
three system sizes viz. $N$ = 100 (uppermost), 200 (middle) and 300 (lowermost). Clearly, the value
of the constant $C$ decreases with increasing system size $N$. See also figure \ref{findconst2}.
}
\label{findconst1}
\end{center}
\end{figure}

\begin{figure}
\begin{center}
\noindent \includegraphics[clip,width= 12cm, angle = 0]
{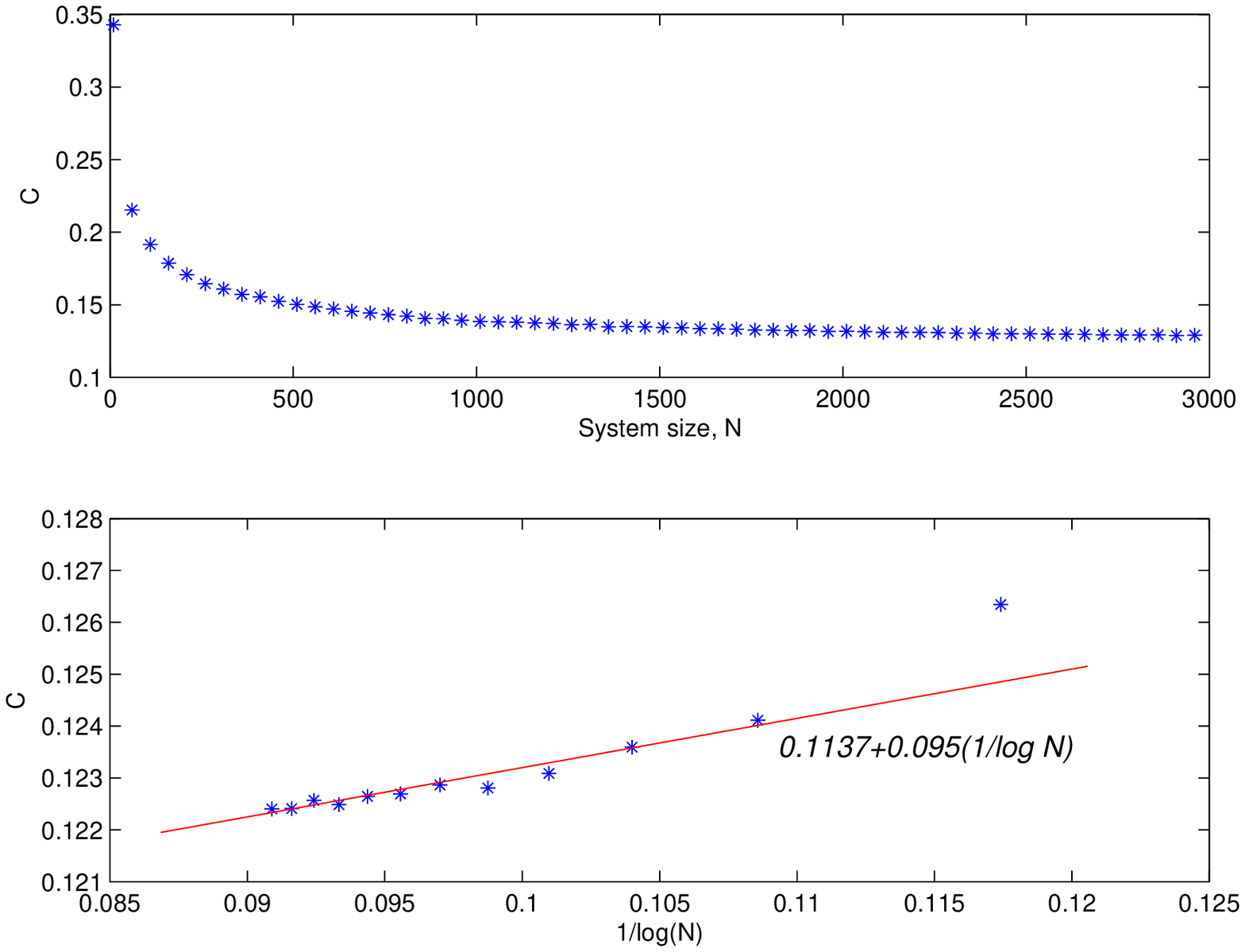}
\caption\protect{{\it Upper panel}: Dependence of $C$ on the system size $N$. 
As we can see the value of $C$ falls rapidly with increasing system size (for small systems). 
With $N=3000$, 
$C\simeq 0.1285$. 
{\it Lower panel}: Dependence of $C$ on $1/log(N)$. 
We simulated systems with different sizes ($N=5000, 10000, 15000, \ldots, 60,000$)
for $\sim 10^5$ time periods. The observation fits
well with $0.1137+0.095/log(N) $. The rightmost point ($N=5000$) is
above the fitted line because of the effect of small system size.
}
\label{findconst2}
\end{center}
\end{figure}

\noindent To solve Eqn. \ref{distrlambda} in the steady state, note that $(1-\lambda_i)E(w_i)=C$, a constant, 
solves the problem. Therefore, we can rewrite the system of equation as (assuming $n=N$)

\begin{eqnarray}
w_1(t+1) &=& \lambda_1 w_1(t) +C\mu_{1(t+1)} \nonumber\\
\ldots       \ldots  \nonumber\\
w_i(t+1) &=& \lambda_i w_i(t) +C\mu_{2(t+1)} \nonumber\\
\ldots        \ldots \nonumber\\
w_N(t+1) &=& \lambda_N w_N(t) +C\mu_{N(t+1)},
\label{distrlambda1}
\end{eqnarray}

\noindent with $\lambda_i\sim$ uniform[0, 1]. Also recall that $\mu_i=N\epsilon_i$ with 
$f(\mu_i)=exp(-\mu_i)$ (see Sec. \ref{subsubsec: ss0}). Ref. \cite{mohanty;06} finds the value of
the constant $C$, in the context of the usual kinetic exchange models with binary trading scheme.
Here, we confirm by simulation that  $(1-\lambda_i)E(w_i)$ is actually a constant 
for all $i$ 
(given the system size
i.e., $N$; see figures \ref{findconst1}  and \ref{findconst2}). Hence,
we can regard Eqn. \ref{distrlambda1} as correctly representing the model (see also 
Ref. \cite{basu-mohanty;08} which treated markets populated with agents each having
a different autoregressive process defining their wealth evolution).
The resultant distribution of the above model is a power law (see figure \ref{zipf}). 
Ref. \cite{asc;11} considered a slightly more general version of this type of maps.
Following Ref. \cite{mohanty;06}, a very simple proof is considered below. Note that (in the steady state) by taking expectations on both sides of Eqn. \ref{distrlambda1}, we can rewrite it as
\begin{equation}
(1 - \lambda_i)E(w_i) = C.
\label{avw}
\end{equation}

\begin{figure}
\begin{center}
\noindent \includegraphics[clip,width= 12cm, angle = 0]
{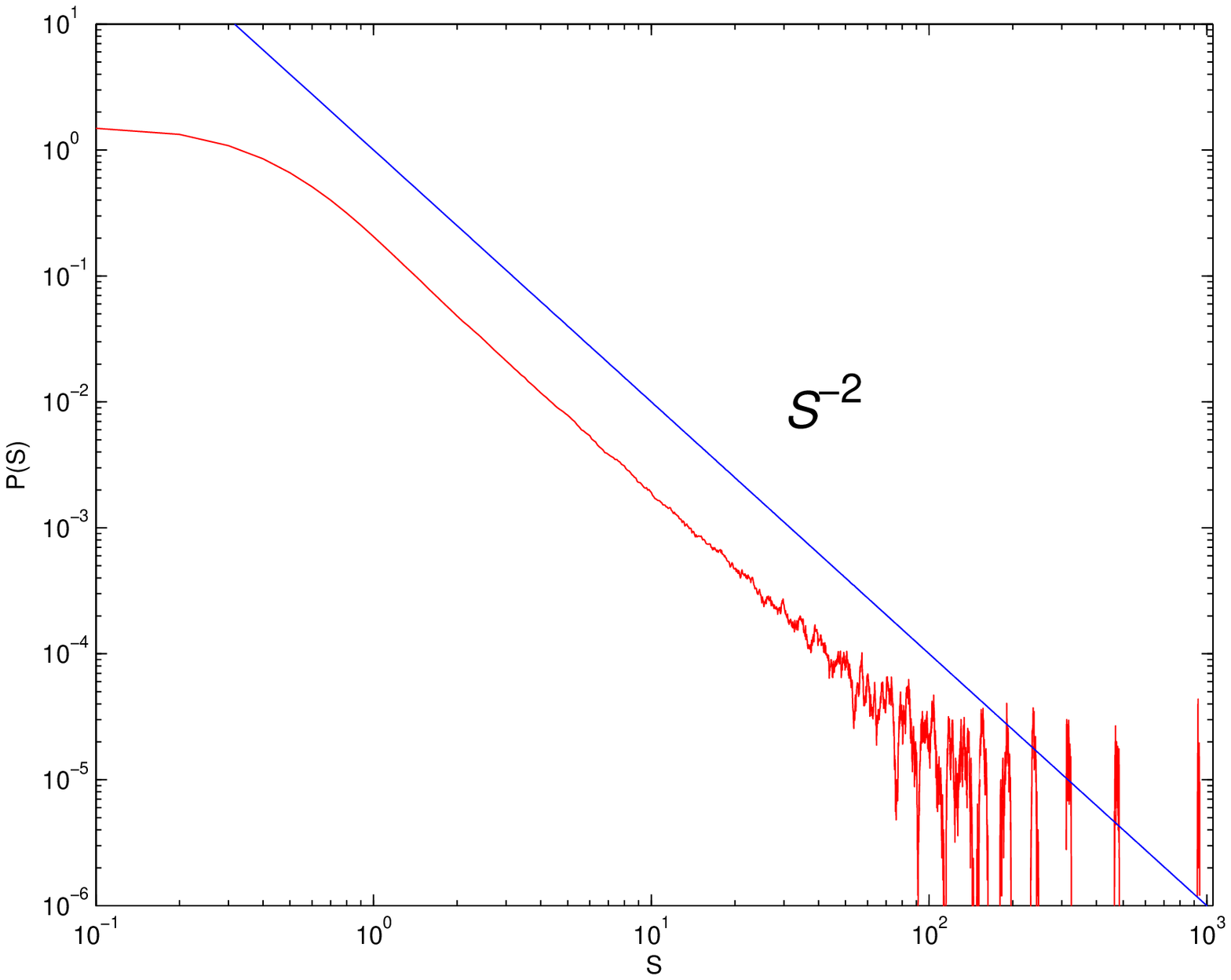}
\caption\protect{Firm size distribution (workers' distribution across the firms): power law
(see also Ref. \cite{axtell;01}). 
All simulations are done
for $O(10^7)$ time steps with 5000 agents and averaged over $O(10^4)$ time steps.
}
\label{zipf}
\end{center}
\end{figure}

\noindent By taking total differentiation and rearranging terms, we get 
$$\frac{d\lambda}{dw}=w^{-2},$$
\noindent where $w$ represents $E(w)$. Hence, the average workforce in a firm with a particular $\lambda$ is given by Eqn. \ref{avw}. Also, the relation between the distribution of $\lambda$ (i.e., $f(\lambda)$) with that of $w$ is given by the following Eqn.
$$P (w)dw = f(\lambda)d\lambda .$$
\noindent The last two equations show that in an array of firms with uniformly distributed $\lambda$, the distribution of $w$ would be
$$P (w)=w^{-2}.$$
\noindent Hence, the firm size has a power law distribution (Zipf's law; see Ref. \cite{axtell;01}).
}

\section{Growth rate} 
\label{sec: growth}
{
\noindent Truly speaking, in this model there is no absolute growth. The economy as a whole is 
completely conserved. There are a few firms (with high $\lambda$) that grows initially. But after
achieving their average size, they do not grow any further in the absolute sense. But of course, fluctuation
is still present. Recall that the reduced form of the model is given by
 
\begin{equation}
w_i(t+1) = \lambda_i w_i(t) +C\mu_{i(t+1)} \hspace{1 cm}\mbox{for all $i$,} 
\label{distrlambda2}
\end{equation}

\noindent which is simply a set of autoregressive type equations. Let us define growth rate as $r_i= w_i(t+1)/w_i(t) $.
Clearly, $r_i=\lambda_i+C\mu_{i(t+1)}/ w_i(t) $ in this model.

\begin{figure}
\begin{center}
\noindent \includegraphics[clip,width= 12cm, angle = 0]
{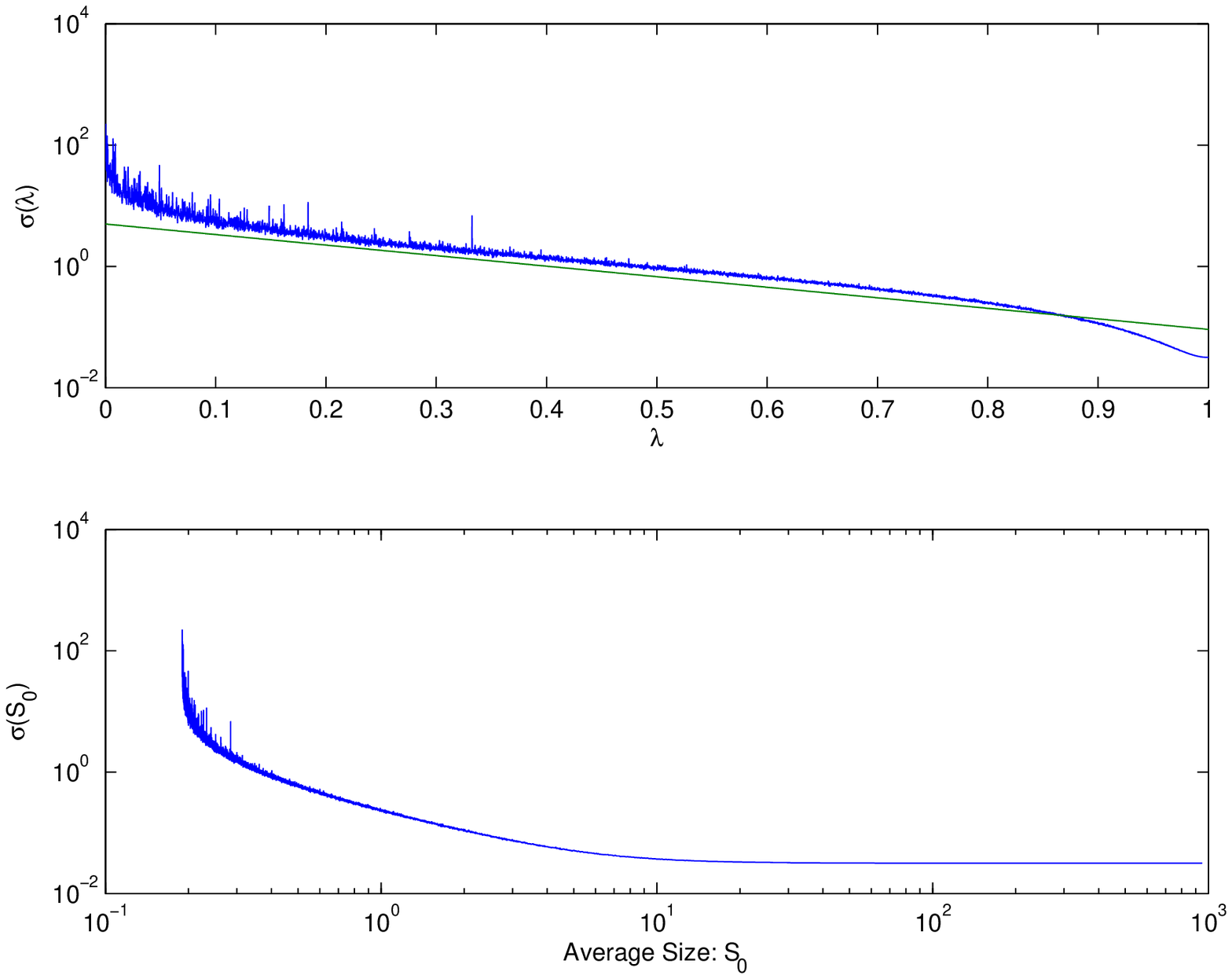}
\caption\protect{Standard deviation of $r_{i(t+1)}=w_i(t+1)/w_i(t)$
decreases with increasing $\lambda$ (upper panel) 
and average size (lower panel). It shows exponential decay with respect to $\lambda$ for 
$0\le \lambda\le 0.9$. The straight line shown in the upper panel is $5.exp(-4\lambda)$.
All simulations are done
for $O(10^7)$ time steps with 5000 agents and averaged over $O(10^4)$ time steps. 
}
\label{stddev}
\end{center}
\end{figure}

\begin{figure}
\begin{center}
\noindent \includegraphics[clip,width= 12cm, angle = 0]
{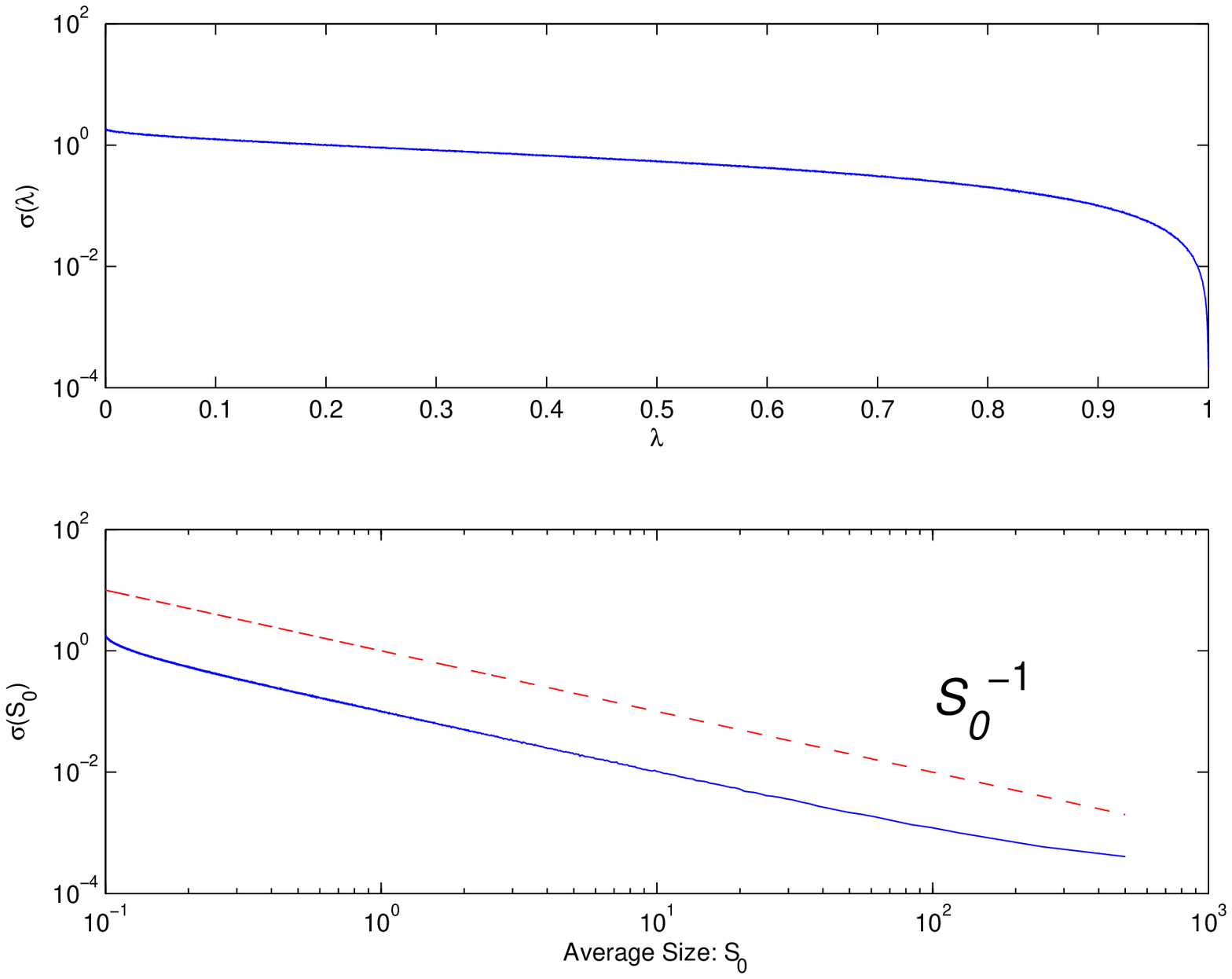}
\caption\protect{Standard deviation of $\log r_{i(t+1)}=\log w_i(t+1)-\log w_i(t)$
decreases with increasing $\lambda$ (upper panel) 
and average size (lower panel). 
Clearly, it shows a power law decay with respect to size as has been documented
in Ref. \cite{stanley;96}. However, the exponent found in this model is -1 whereas
data suggests that it is 0.16 $\pm$ 0.03.
All simulations are done
for $O(10^7)$ time steps with 5000 agents and averaged over $O(10^4)$ time steps. 
}
\label{stddevlog}
\end{center}
\end{figure}

\begin{figure}
\begin{center}
\noindent \includegraphics[clip,width= 12cm, angle = 0]
{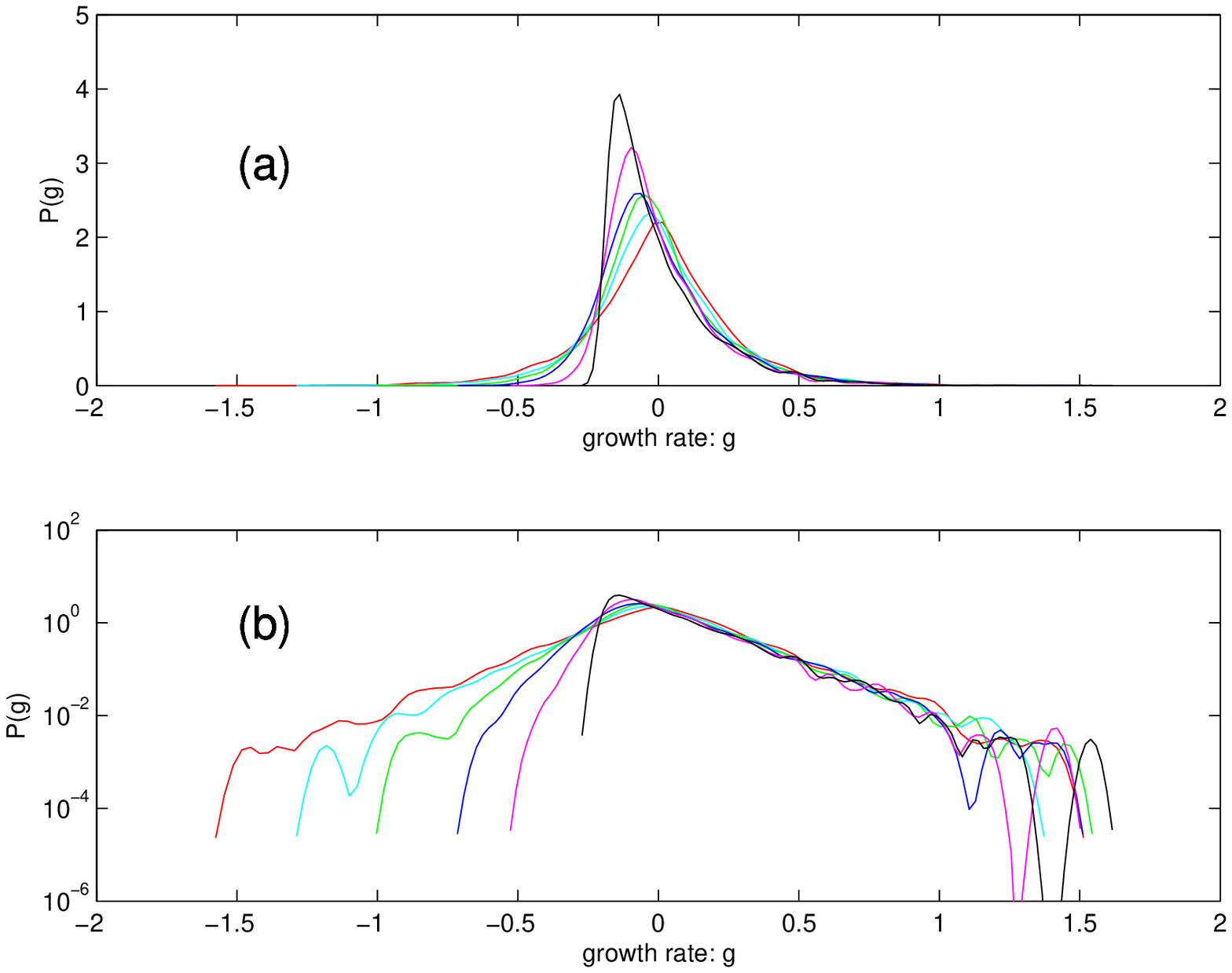}
\caption\protect{Distribution of growth rates $g_i(t)=w_i(t)-w_i(t-1)$ for turnover rates 
$\lambda_i$ = 0, 0.2, 0.4, 0.6, 0.8, 0.95. 
Evidently for small $\lambda_i$, $g_i$ has Laplace distribution (bi-exponential). As $\lim$ 
$\lambda\rightarrow 1$, the distribution becomes one sided exponential only 
(see Eqn. \ref{distrlambda2} and note that the error term is exponentially distributed). 
However, contrary to what we get here, 
the data suggests that $\log w_i(t)-\log w_i(t-1)$ has a Laplace distribution (Ref. \cite{stanley;96}).
All simulations are done
for $O(10^7)$ time steps with 5000 agents and averaged over $O(10^4)$ time steps. 
}
\label{growthrate}
\end{center}
\end{figure}

\noindent Evidently, as $\lambda_i$ rises, the average size $E(w_i)$ also rises. Therefore, the variance
(or the standard deviation) of the growth rate falls with increasing size (see figure \ref{stddev}). 
This model captures this feature well though it does not match the exact exponent. 
The standard deviation 
decreases following a power law with exponent -1 (see figure \ref{stddevlog})
whereas the real data set suggests that the exponent is 0.16 $\pm$ 0.03 (see Ref. \cite{stanley;96}). 
We also studied the distribution of the growth rate of the firms. Ref. \cite{stanley;96} defined growth-rate
as $\log r$ with $r$ defined as above. However, we can approximate the growth rate as following
$\log r_{it}=\log (w_i(t)/w_i(t-1))\simeq (w_i(t)-w_i(t-1))/w_i(t)$ (by adding and subtracting 1 to $r$).
Note that for $\lambda=0$, the distribution of any firm (i.e., $w(t)$) would be exponential. Therefore,
the numerator in the expression of $\log r_t$ has a Laplace distribution (see figure \ref{growthrate}). 
But the growth-rate $\log r$ as has been defined in \cite{stanley;96} clearly does not have a Laplace distribution in this model. In fact, in no way it resembles the proposed Laplace distribution (it has too many discrete jumps, specially for firms with small $\lambda$ ). 
Hence, our model does not perform well to reproduce the pdf of growth rate found in
Ref. \cite{stanley;96}.
 
}

\section{Summary} 
\label{sec: summary}
{
\noindent We have studied a model of firm dynamics. There are a number of well known 
statistical regularities in the firm dynamics (see for example Ref. \cite{axtell;01, stanley;96}).
The main features considered here are (a) power law decay in size distribution, (b) reduction in fluctuation
in growth rate with increasing size of the firm (following another power law) and 
(c) Laplace distribution of the growth rate. 
Ref. \cite{axtell;01} mentions another regularity concerning the distribution of 
payments to the workers in different firms. But we have neglected it completely
because we did not consider any strategic behavior on the part of the workers or the firms (i.e.,
the firm owners) . 

\noindent The model that we propose is a multi-agent model with 
$n$-ary interactions ($2\le n\le N$) at each time step. We  
present some analytical results on the steady state distributions for constant
turnover rate (across the firms)
obtained from 
the model which seems to not agree with the earlier approximate results (see Ref. \cite{chakraborti;04}).
Specifically, we show that if the number of interacting firms goes to infinity the
exact distribution of the firm size is given by an infinite sum of weighted exponentials. Hence,
at least in this limit the solution is definitely not gamma distribution as had been proposed
for binary interaction.  With distributed turnover rates, the model can very easily produce the power law
in size distribution as is done by the usual kinetic exchange models with binary interactions.
However, the importance of the generalization from binary to n-ary interaction lies
in the idea that it better captures the workers' flow among a huge number of firms.
Next, we study the distributions of the growth rates of the firms. We show that
this model quantitatively captures the observation that
the fluctuations in growth rates fall according to a power law with increasing firm sizes, though 
it fails to match the exact
exponent (the model gives 1 whereas the data suggests 1/6). 
Another shortcoming of the model is that, in this model, all
growths are relative i.e., the economy as a whole is not growing which is certainly
not the case with the real economies.
The major point of departure of our model from {\it The law of proportionate effect} (see Ref. \cite{gibrat}
which assumes zero effect of the firm size on its growth rate)
is that we assume autocorrelation in the growth rate. In fact, we present the firm's growth process by
an autoregressive process of order one so that the growth rate is affected by the size. See Ref. \cite{autocorr}
which documents autocorrelations in the firm growth processes. Ref. \cite{coad} claims that small firms actually show a negative autocorrelation whereas the large firms have positive autocorrelation. 
However, the evidence is not conclusive.   

\noindent In the appendix, we discuss briefly about how this generalized version of the
kinetic exchange models
are related to the generalized Lotka-Volterra model and also, how can we
apply the model stated above to model the income/wealth
distributions.

}
\noindent {\bf Acknowledgement}: The author is grateful to Bikas K. Chakrabarti and Arnab Chatterjee 
For some useful comments.

\section{Appendix}

\subsection{Comparison with the generalized Lotka-Volterra  (GLV) model } 
\label{sec:GLV}
{
\noindent Ref. \cite{richmond;arxiv} raised a question that whether and how, the kinetic exchange models are related
to the generalized Lotka-Volterra model (both can produce power law distributions). One major obstacle
in answering the question was the problem that the kinetic exchange models focused on
binary interaction and GLV takes in to account all the agents in each interaction. However, given the
above formulation, one can directly compare the two mechanisms.

Ref. \cite{solomon} presents the GLV mechanism by the following equation (for $1\le i\le N$),
\begin{equation}
w_i(t+1)=\lambda(t+1)w_i(t)+a(t)\bar{w}(t)-c(t)w_i(t)\bar{w}(t).
\end{equation}
\noindent Two notable differences of GLV with the model proposed above are the presence of 
a time varying $\lambda$ and a nonlinear interaction via the average $w$ ($\bar{w}$) in the GLV.
 Recall that in the model proposed as a generalized kinetic exchange model, the average is always
fixed (at unity, in this case). Ref. \cite{solomon} reduces the system to $N$ decoupled equation of the
following form,
\begin{equation}
v_i(t+1)=\lambda(t)v_i(t)+a(t).
\label{GLV}
\end{equation}

\noindent Comparing Eqn. \ref{GLV} with Eqn. \ref{distrlambda2}, we see that the essential difference
between the two systems is whether $\lambda$ varies over time or not. Another important point is that in GLV, $\lambda$ has to be greater than 1 sometimes which is not possible in the other case. 
In short, we can say that the GLV
mechanism depends on the process of random multiplicative maps whereas the generalized kinetic exchange model does not.  

}

\subsection{Generalized exchange model as an equilibrium outcome} 
\label{subsec: genex}
{
\noindent In this subsection, we discuss how the model with $N$-ary interaction be applied to
model wealth/income distribution. Though we think that binary trading is much more common
in the real market place (than $N$-ary trading), we present briefly a direct generalization of the framework presented in \cite{chakrabartis;09} which dealt with binary trading mechanism. 

\medskip
 
\noindent Let there be $N$ agents each having 1 unit of perfectly divisible money in their possession
(at the beginning of all trading). At each
period each of them produces $Q$ unit of commodities ($Q_i$ may be different from $Q_j$ for 
all $i$ and $j$) such that no two commodities are the same. 
Let the preference of the $i$-th agent be defined as 
$$ U_i=x_1^{\alpha_1}x_2^{\alpha_2}   \ldots x_N^{\alpha_N}         m_i^{\alpha_m} .  $$
The budget constraint would be $p_1x_1+p_2x_2+ \ldots +m_i\leq M_i+p_iQ_i$. We make 
the standard assumption that $\alpha_1+\alpha_2+\ldots +\lambda=1$ where $\lambda=\alpha_m$ is the savings propensity. Then we can write the
constrained maximization problem as
\begin{equation*}
{\mathcal L}=x_1^{\alpha_1}x_2^{\alpha_2} \ldots  
m_1^{\lambda} -\bar{\mu}(p_1x_1+p_2x_2+\ldots m_i- M_i+p_iQ_i) 
\end{equation*}
\noindent where $\bar{\mu}$ is the Lagrange multiplier. Solving the optimality conditions, we get (denoting $\alpha_m$ by $\lambda$)
$$m^*_i=\lambda(M_i+p_iQ_i)$$
\noindent for all $i$.
Solving for equilibrium price vector, one derives
\begin{equation*}
m_i(t+1) =\lambda m_i(t) + \epsilon_i(1-\lambda)\sum_j m_j(t) 
\end{equation*}
\noindent where $\epsilon_j$ can be suitably defined as a beta $r. v.$.
Clearly, the above equation reduces to the following
\begin{equation*}
m_i(t+1)=\lambda m_i(t) + \mu_{i(t+1)}
\end{equation*}
\noindent (this is exactly the system we studied above in section \ref{sec: basic}).
}

\end{document}